\documentclass{PoS}

\title{Multitrace matrix models of fuzzy field theories}

\ShortTitle{Multitrace matrix models of fuzzy field theories}

\author{\speaker{M\'aria \v{S}ubjakov\'a}, Juraj Tekel \\
        Department of Theoretical Physics \\
        Faculty of mathematics, physics and informatics \\
        Comenius University, Bratislava, Slovakia\\
        E-mail: \email{maria.subjakova@fmph.uniba.sk, juraj.tekel@fmph.uniba.sk}}

\abstract{We review analytical approaches to scalar field theory on fuzzy spaces. We briefly outline the matrix description of these theories and describe  various approximations to the relevant matrix model. We discuss the challenge  of obtaining  a consistent approximation that includes the higher moments of the theory.  }

\FullConference{Corfu Summer Institute 2019 "School and Workshops on Elementary Particle Physics and Gravity" (CORFU2019)\\
		31 August - 25 September 2019\\
		Corf\`u, Greece}
		
\usepackage{amsmath}

\begin{document}

\section{Introduction}

Field theories defined on fuzzy spaces provide an important setting to the study of the consequences of non-commutativity of the underlining space. With their finite degrees of freedom, fuzzy spaces are readily accessible by numerical computation tools. Therefore, they have been extensively studied by these methods \cite{samo numerics, numerika sphere 2, numerika sphere 3,numerika disk 1, numerika time sphere, numerika torus, numerika ? 1, numerika ? 2}, showing very different properties than their commutative counterparts even in the commutative limit.

Fuzzy field theories have been studied also analytically \cite{numerika ? 3, analytika 1, analytika 2, analytika 3, analytika 4, analytika 5, analytika 6, analytika 7, non perturbative, juro 1, juro 2, juro 3, analytika 8}. However, the corresponding matrix models are complicated, and their full solution has not been obtained yet. Therefore, various approximations of the model have been considered.

In this paper, we review such analytical approaches mainly for the fuzzy sphere. However, the same analysis can be used for other fuzzy spaces, e.g. higher fuzzy $\mathbb{ C}P^n$ spaces. 

We start with a brief overview of the matrix model corresponding to the scalar field theory on the fuzzy sphere. Afterwards, we discuss the non-perturbative second model approximation of the model and compare it to the numerical simulations of the full model. In the last section, we review the perturbative approximation that considers also few higher moments of the theory, and demonstrate the importance of inclusion of these higher moments in a non-perturbative way.

\section{Scalar fuzzy field theories}
Fuzzy spaces are compact non-commutative spaces. The algebra of functions defined on such spaces has a finite number of degrees of freedom and can be defined as some matrix algebra \cite{fuzzy spaces 1, fuzzy spaces 2, fuzzy spaces 3,fuzzy sphere 2}. 

Scalar field theories on fuzzy spaces can be given in terms of the correlation functions:
\begin{equation}
\langle O[M] \rangle = \frac{1}{Z} \int dM e^{-S[M]}O[M] \label{Z}
\end{equation}
and therefore correspond to random matrix models with probability measure given by the action of the theory.

One of the simplest example of fuzzy spaces is fuzzy sphere \cite{fuzzy sphere 1}. The algebra of functions on the fuzzy sphere is spawned by the ${N \times N}$ matrices:

\begin{equation}
X_i = \frac{2R}{\sqrt{N^2-1}}L_i, 
\end{equation}

where $L_i$ are ${su(2)}$ algebra generators and $R$ gives the sphere radius. It is straightforward to check that these matrices satisfy the non-commutative relation:
\begin{equation}
[X_i,X_j] = i \frac{2R}{\sqrt{N^2-1}} \epsilon_{ijk}X_k, \qquad \sum_{i=1}^3 X_iX_i =R^2. \label{fuzzy sphere}
\end{equation}
We can verify that the large $N$ limit corresponds to the commutative limit as we recover  algebra of the ordinary commutative sphere functions  in (\ref{fuzzy sphere}).

The real scalar field theory on the fuzzy sphere is given as hermitian random matrix model (\ref{Z}) with the action:
\begin{equation}
S[M]= Tr \bigg (\frac{1}{2} m^2 M^2+ gM^4+ \frac{1}{2} M \mathcal{K}M \bigg), \label{action}
\end{equation}
The kinetic term in the action (\ref{action}) corresponds to the ${su(2)}$ quadratic Casimir operator:
\begin{equation}
\mathcal{K}M = [L_i,[L_i,M]].
\end{equation}
We consider only the quartic theory with the interaction parameter $g$.
 
To treat such models analytically in the large $N$ limit we need to rewrite the integral (\ref{Z}) in terms of different integration parameters- the eigenvalues of the matrices and the remaining angular degrees of freedoms. This leads to a model:
\begin{equation}
Z= \int \bigg( \prod_{i=1}^N d \lambda_i \bigg) e^{-N^2\big[\frac{1}{2}m^2\frac{1}{N}\sum \lambda_i^2+ g\frac{1}{N}\sum \lambda_i^4-\frac{2}{N^2}\sum_{i< j} \log|\lambda_i-\lambda_j|\big]} \int dU e^{-\frac{1}{2}Tr[U^{\dagger}\Lambda U\mathcal{K}(U^{\dagger}\Lambda U)]}, \label{model}
\end{equation}
where ${\lambda_i}$ denotes the matrix eigenvalues, ${\Lambda= \textrm{diag}(\lambda_1,\lambda_2, \ldots, \lambda_N)}$ and ${U \in U(N)}$. 

The kinetic term is not invariant under unitary transformations. Therefore the angular integral in (\ref{model}) is non-trivial and  we are unable to perform the integration analytically.  Thus the kinetic contribution in the action presents a significant obstacle in the analytical attempts to obtain the model solution.

As the full analytical solution of the model (\ref{model}) cannot be attained, the approximations of the model have been considered. In the next sections we review some of these approximative techniques. 
 
\section{Second moment approximation}

For the following sections it is useful to define the notion of the effective action as
\begin{equation}
e^{-N^2S_{eff}[\Lambda]}=  \int dU e^{-\frac{1}{2}Tr[U^{\dagger}\Lambda U\mathcal{K}(U^{\dagger}\Lambda U)]}. \label{effective action}
\end{equation}

The effective action depends only on the eigenvalues or equivalently, due to the kinetic term invariance under translation ${M \rightarrow M+a\mathbb{I}} $, on the symmetrized moments of the theory:
\begin{equation}
t_n = Tr \bigg (M -\frac{1}{N}(TrM)\mathbb{I} \bigg)^n.
\end{equation}

The analytical solution that can be obtained in the case of the free theory, i.e. for ${g=0}$ \cite{analytika 6, analytika 7}, provides us with the possibility to determine  a part of the effective action that depends on the second symmetrized moment non-perturbatively \cite{non perturbative}. The effective action then can be rewritten in the form:
\begin{equation}
S_{eff}= \frac{1}{2}F(t_2)+ R,
\end{equation}
where the function $F(t_2)$ captures the known results in the case of the free theory and the remaining part $R$ gives zero contribution for such free model. For the fuzzy sphere the second moment function comes out as:
\begin{equation}
F(t_2)= \log \bigg (\frac{t_2}{1-e^{-t_2}} \bigg). \label{second moment model}
\end{equation}

We can then approximate the effective action with this function and drop the remaining term as an approximation. Such matrix models that depend only on the eigenvalues can be studied in the commutative limit using the saddle point method \cite{juro 2, fuzzy spaces 1, fuzzy spaces 2, random 1, random 2}. This method tells us that in the commutative limit ${N\rightarrow \infty}$,  only the most probable eigenvalue distribution contributes to the integral (\ref{model}). This is the configuration satisfying:
\begin{equation}
\frac{\partial S}{\partial \lambda_i} =0. \label{saddle}
\end{equation}

For fuzzy field theories, this saddle point equation gives three types of solutions depending on the values of parameters in the action $m^2$, $g$. Those three types of solutions are referred to as:

\begin{itemize}

\item the disorder phase - where the matrix eigenvalues are distributed symmetrically around zero, over one continuous interval, 

\item the uniform order phase- in which  the eigenvalues are distributed around one of the minima of the potential ${V(\lambda) = \frac{1}{2}m^2\lambda^2 +g \lambda^4}$ for the negative values of $m^2$, 

\item the non-uniform order phase- in this case, the eigenvalue distribution is not supported over one continuous interval, but half of the eigenvalues lay around one of the potential minima and a half around the other one.
\end{itemize}

The last phase does not exist in commutative field theories, it is typical for fuzzy field theories \cite{numerika ? 3, com phases 1, com phases 2, com phases 3, com phases 4}. This phase is non-local as it corresponds to the field oscillating around different potential minima in different parts of space and is regarded as a consequence of so-called UV/IR mixing \cite{uv/ir 1, uv/ir 2}.

 This is the scale mixing phenomenon that plagues fuzzy field theories, as  one cannot separate the processes on small and large scales.  The more we localise the event in one dimension, the more de-localised it became in the other ones.   The UV/IR mixing does not disappear in the commutative limit. Therefore the non-uniform phase remains also present.

 More than one solution may be possible for some values of the parameters $m^2$, $g$. In such case the solution  with lower free energy is realized:
\begin{equation}
\mathcal{F}= -\frac{1}{N^2}\log \bigg (\int dM e^{-N^2S[M]} \bigg). \label{free energy}
\end{equation}

\begin{figure}
  \centering
  \includegraphics[width=0.6\linewidth]{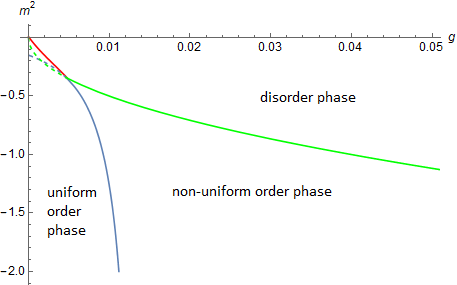}
  \caption{The phase diagram of the second moment approximation (\ref{second moment model}), obtained in \cite{juro 3}. The green line  denotes the transition  line between the disorder phase and the non-uniform order phase, the red line the transition between the disorder phase and the uniform order phase. The transition line between the uniform order and the non-uniform phase is pictured in blue. }
  \label{figure 21 minimovka2 }
\end{figure}

The saddle point equation was solved for the fuzzy sphere numerically in \cite{juro 1}. The perturbative solution in large negative $m^2$ parameter was obtained in \cite{juro 3}. The fact that though the perturbative expansions itself do not converge, the Padé approximations of  the series give reasonable results in agreement with the numerical solution was used. The obtained results are pictured in Figure \ref{figure 21 minimovka2 }.

 We can see that three transition lines meet at the triple point of the theory, as is expected from the numerical simulations of the full model (\ref{model}). The location of the triple point was determined as ${g_c = 0.0048655}$. These results are in a reasonable agreement with the most recent numerical simulation \cite{samo numerics}. Thus the second moment approximation works reasonably well around the origin of the parameter space. 

However, it gives qualitatively very different results further from the origin. As we can see in Figure \ref{figure 21 minimovka2 }, the transition line between the uniform order and non-uniform order phase asymptotically approaches a finite value ${g = \frac{1}{16e^{3/2}}}$. Therefore, the uniform order phase is  not realized at all for the larger values of the interaction  parameter $g$. This is in disagreement with the numerical simulations which suggest that the uniform order phase reaches all the values of $g$ and the transition line behaves linearly.  

These differences are the consequence of the considered approximation, and we need to include also the higher moments of the theory in order to study the model further from the origin in the parameter space.

\section{Beyond second moment approximation}

As we have discussed in the previous section, the second moment approximation does not capture well all of the features of the full model. Mainly, it comes short further from the origin of the parameter space. Thus, the inclusion of  higher moments is needed. 

An approximation that considers also higher moments in some way comes from the perturbative expansion of the unitary integral in (\ref{model}) \cite{analytika 1, analytika 2}. This expansion was obtained up to the fourth order and leads to the following multitrace expression for the effective action \cite{analytika 3}:

\begin{equation}
S_{eff}=  \frac{1}{2}\bigg(\frac{1}{2}t_2-\frac{1}{24}t_2^2+\frac{1}{2880}t_2^4 \bigg )-\frac{1}{432}t_3^2- 
\frac{1}{3456} \left(t_4-2t_2^2 \right)^2 . \label{multitrace}
\end{equation} 

Note that in the previous section, we reviewed the approximation that disregarded all higher moments completely,  but the terms depending on the second moment were considered non-perturbatively. Now we include also a few higher moments, however, we only take the first terms of the perturbative expansions as the full functions are unknown.

\begin{figure}
  \centering
  \includegraphics[width=0.6\linewidth]{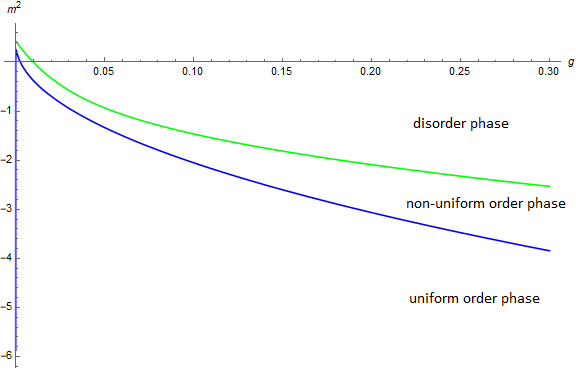}
  \caption{The phase structure of the model (\ref{multitrace}), obtained in \cite{juro 2}. The green line denotes the transition between the disorder and the non-uniform order phase, the blue line pictures the transition between the non-uniform order and uniform order phase. }
  \label{figure 2 perturbative }
\end{figure}

The model with the effective action (\ref{multitrace}) was solved numerically in \cite{juro 2} and led to the phase structure shown in Figure \ref{figure 2 perturbative }.

We see that this approximation completely fails to reproduce the phase structure near the origin of the parameter phase. The phase transition lines do not intersect and there is no triple point. Moreover, the transition lines do not cross the origin of the parameter space, and the phase transition  between the uniform order and the non-uniform order  does not behave linearly. 

Despite this, the approximation has some relevance further from the origin, particularly for the disorder phase, as  the moments of the solutions are small for the large interaction parameter $g$.

So we see that the multitrace expansion of the effective action is not a sufficient tool to study the phase structure of the fuzzy field theories and
we need to consider higher moments non-perturbatively. These non-perturbative functions are not known, so it seems like we cannot do much more. However, it could be useful to study what kinds of changes to the phase structure we can achieve by including these higher moments non-perturbatively. We can, therefore, consider various functions of higher moments and study their effect on the phase diagram.

The higher moment functions cannot be completely arbitrary. In addition to the first known orders of their small expansions (\ref{multitrace}), the properties of the full model (\ref{model}), known from the numerical simulation,  present some other restrictions. For example, following functions of the higher moments could be considered:
\begin{equation}
F_3  \left( t_3 \right) = a \log \left(1+t_3^2 \right) \qquad \text{or} \qquad F_3  \left( t_3 \right) =  \log \left(1+b\,t_3^2 \right)
\end{equation}
and equivalently for the function of the fourth moment $F_4\left( t_4-2t_2^2 \right)$, with the coefficients $a,b$ set to match the multitrace expansion (\ref{multitrace}). 

To study such general functions, analytical tools are necessary. The analytical perturbative method that was used to solve the second moment approximation model in \cite{juro 3} can be generalised also to the higher moments. The saddle point equations can be solved perturbatively in the large negative parameter $m^2$ also when considering the functions of the third and fourth moments with well behaved small and large expansions. Therefore, this method  provides us with a possibility to study exactly such higher moments models.

However, this generalisation is not completely straightforward, especially for such functions of higher moments that increase at higher than the logarithmic rate for the large values of their parameters. Moreover, obtaining the transition  line between the disorder phase and the uniform order phase is technically difficult even in the second moment approximation, due to the fact that the disorder phase does not exist in the large negative $m^2$ limit. In the case of the second moment approximation this problem is bypassed by expanding the solutions around the transition between the disorder and the non-uniform order phase, which is known exactly. However, for the general effective action that depends also on higher moments, this is not always the case.

Nevertheless, we believe that these problems can be handled and we are currently working on the consistent inclusion of the higher moments to the effective action.

\section{Conclusions and outlook}

We reviewed two main approaches to the analytical treatment of the matrix models corresponding to the fuzzy scalar field theories.  The first approach considered only the second moment of the theory but did it in a non-perturbative way.  This model gave good results near the origin of the parameter space, but further away it was inconsistent with the numerical simulations.

The second approach also considered the third and forth moments of the theory but only  in a perturbative way. This model did not reproduce the phase structure of fuzzy field theory at all and, therefore,  highlighted the importance of including the higher moments in a non-perturbative way.

However, these higher moments terms are known only perturbatively.
Therefore, we outlined the possibility to study the general effective action depending also on these higher moments. This option arose recently with development of the analytical perturbative tools to solve the saddle point condition (\ref{saddle}) for the second moment approximation. However, the road to the complete generalisation of this approach to the higher moment approximation is yet long.

\end{document}